\documentstyle[12pt]{article}
\newcommand{\bfr}{\begin{flushright}}
\newcommand{\efr}{\end{flushright}}
\newcommand{\bfl}{\begin{flushleft}}
\newcommand{\efl}{\end{flushleft}}
\begin{document}
\vspace*{-2cm}
\bfr{KNCT-HEP-9601}\efr\vspace{-9mm}
\bfr{OTSUMA-HEP-9601}\efr\vspace{-9mm}
\bfr{April 1996}\efr
\vspace{1cm}
\begin{center}
{\Large \bf 2+1 dimensional charged black hole}\\
\vspace{5 mm}
{\Large \bf with}\\
\vspace{5 mm}
{\Large \bf (anti-)self dual Maxwell fields}\\
\vspace{2cm}
Masaru Kamata
\footnote[1]
{e-mail address: kamata@gokumi.j.kisarazu.ac.jp} \\
  \vspace{0.7cm}
  {\it Kisarazu National College of Technology\\
    2-11-1 Kiyomidai-Higashi, Kisarazu, Chiba 292, Japan}\\
  \vspace{1cm}
Takao Koikawa
\footnote[2]
{e-mail address: koikawa@csc.otsuma.ac.jp}\\
  \vspace{0.7cm}
{\it School of Social Information Studies, Otsuma Women's University\\
Karakida, Tama, Tokyo 206, Japan}\\
  \vspace{1.5cm}
  ABSTRACT\\
\end{center}
  \vspace{.5cm}
\hspace{1cm} We discuss the exact electrically charged BTZ black hole 
solutions to the Einstein-Maxwell equations with a negative cosmological 
constant in 2+1 spacetime dimensions assuming a (anti-)self dual condition 
between the electromagnetic fields. In a coordinate condition there appears 
a logarithmic divergence in the angular momentum at spatial infinity. We 
show how it is to be regularized by taking the contribution from the 
boundary into account. We show another coordinate condition which leads to a 
finite angular momentum though it brings about a peculiar spacetime 
topology.
\vspace{.5cm}

\bfl
keywords: black hole, $ 2+1 $ dimensions, self dual
\efl
\thispagestyle{empty}
\newpage
\setcounter{page}{2}
The (2+1)-dimensional black hole solution with a negative cosmological 
constant, which corresponds to the Kerr black hole solution in 
(3+1)-dimensional general relativity, was found by Ban\~ados, Teitelboim and 
Zanelli (BTZ) \cite{BTZ,BHTZ}. It has mass and angular momentum. Then an 
exact electrically charged BTZ solution to the Einstein-Maxwell equations 
with a negative cosmological constant in 2+1 spacetime dimensions assuming a 
self dual(SD) (anti-self dual(ASD)) equation was discussed 
\cite{KK,Clement}. It is the solution with mass $M$ and angular momentum $J$ 
which were determined by the asymptotic form of the metric. However the 
angular momentum has a logarithmic divergence at spatial infinity and so the 
solution should not be called the black hole solution \cite{Chan}.

In this paper we clarify the origin of the divergence and how we can 
circumvent the problem. In order to discuss the conserved charges we should 
pay attention to the definitions of them. When there is nonzero cosmological 
constant, we can not justify the naive definition of mass and angular 
momentum because they depend on the normalization of Killing vector fields, 
which should be contrasted with the asymptotically flat case with zero 
cosmological constant. We should take the contribution from the boundary 
into account and, in this case, we have an arbitrariness in defining the 
gravitational actions which leave the classical equations of motion 
unchanged.  A class of gravitational actions which is a functional of the 
metric on the boundary does not change the classical equations of motion 
even if they are added to the original action. Though this does not affect 
the equations of motion, this does bring about an arbitrariness of the 
energy and angular momentum in their definitions. We follow the formulation 
of quasilocal energy and conserved charges in Refs.\cite{York,BY,BCM}. Under 
the definitions of mass and angular momentum in the asymptotically anti-de 
Sitter spacetime, they are not constants but functions of positions in 
general. In our previous paper \cite{KK}, we showed a general solution to 
the Einstein-Maxwell equations with a negative cosomological constant with 
one of the coordinate conditions by which all the metric components are 
expressed in terms of  the radial coordinate variable $r$. This turns out to 
define the perimeter length of a circle with $t=$const. and $r$=const. The 
coordinate condition thus determines the spacetime geometry.  In our 
previous solution there is a horizon, but the coordinate conditions which we 
adopted there lead to the divergence of an angular momentum  both at the 
horizon and at the  infinity. Therefore it is not a black hole solution. 
However this does not mean that the general solutions with SD(ASD) 
electromagnetic fields always have the divergence somewhere on or outside of 
the horizon, because we showed only one of infinite possible coordinate 
conditions. In our previous paper we neglected the contribution from the 
boundary which does not change the classical equations of motion but can 
contribute to the energy and charges in general. We should investigate these 
possibilities.
   The present paper is constructed as follows. We start with the summary of 
our general solution. We then define the definitions of energy, mass and 
angular momentum following Refs.\cite{York,BY,BCM}. We discuss some of the 
possible coordinate conditions. In the coordinate condition we adopted in 
the previous paper there appeared a logarithmic divergence in the angular 
momentum. We discuss the regularization of the divergence by the counter 
term emerging from the action defined on the boundary. We then discuss other 
coordinate conditions which suppress the divergence of the angular momentum 
even when $(j_0)_{\phi}$ in \cite{York,BY,BCM} is taken to be zero.

   We shall obtain the solutions by assuming the axisymmetry with nonzero 
electric and magnetic fields between which we assume  a self dual(SD) 
(anti-self dual(ASD)) relation. The Einstein-Maxwell action is
\begin{equation}
     	S= {1 \over{16\pi G}}\int \sqrt{-g}(R-2\Lambda-4\pi GF^2)d^3x -S^0,
     	\label{eq:Action}
\end{equation}
where $G$ is Newton's constant, $\Lambda$ is the negative cosmological 
constant and $F^2 \equiv \linebreak g^{\mu \nu}g^{\rho \sigma}F_{\mu \rho}
F_{\nu \sigma}$.  Here $S^0$ is an action which is a functional of the 
metric and their derivatives defined on the boundary.  The Einstein equation 
is given by
\begin{equation}
	G_{\mu \nu}+\Lambda g_{\mu \nu}=8\pi GT_{\mu \nu},
	\label{eq:Ein}
\end{equation}
where $T_{\mu \nu}$ is the energy-momentum tensor of the electromagnetic 
field:
\begin{equation}
	T_{\mu \nu}=F_{\mu \rho}F_{\nu \sigma}g^{\rho \sigma}-{1\over 4}g_{\mu 
\nu}F^2.
	\label{eq:EMtensor}
\end{equation}
The Maxwell equation is given by
\begin{equation}
	\partial_\rho(\sqrt{-g} g^{\mu \nu}g^{\rho\sigma}F_{\nu \sigma})=0.
     	\label{eq:Max}
\end{equation}
 We use the line element
given by
\begin{equation}
	ds^2=-N^2dt^2+L^{-2}dr^2+K^2(N^{\phi}dt+d\phi)^2,
\label{eq:Metric}
\end{equation}
where $N$, $L$, $K$ and $N^\phi$ are functions of r only.

 Since we have two Killing vectors $\partial/\partial t$ and 
$\partial/\partial\phi$ in the stationary axisymmetric spacetime, we can 
solve the Maxwell equation (\ref{eq:Max}) as follows:
\begin{eqnarray}
	E_{\hat r} \equiv F_{\hat t \hat r}&=&{{C_1}\over K},
	\label{eq:Elec}\\
	{B_{\hat{~}}} \equiv F_{\hat r \hat {\phi}}&=&{{C_1 N^{\phi}+C_2}\over 
N},
	\label{eq:Mag}
\end{eqnarray}
where $C_1$ and  $C_2$ are constants of integration. We impose the SD(ASD) 
equation on the electric and magnetic fields:
\begin{equation}
	E_{\hat r}=\varepsilon B_{\hat{~} },
	\label{eq:SD}
\end{equation}
where $\varepsilon$ takes 1 or -1 corresponding to SD or ASD equation, 
respectively.

As far as we do not have the asymptotically flat spacetime, we can not 
define mass and angular momentum as those observed at spatial infinity. We 
follow the discussion of qusilocal energy and conserved charges in 
Refs.\cite{York,BY,BCM}. The energy $E$, angular momentum $J$ and mass $M$ 
are expressed in terms of the metric components as
\begin{eqnarray}
	E&=&-2LK'-2 \pi K {\varepsilon}_0,
	\label{eq:Energy}\\
	J&=&\displaystyle\frac{K^3L(N^{\phi})'}{N}-2 \pi K (j_0)_{\phi},
	\label{eq:Angmom}\\
	M&=&EN-JN^{\phi},
	\label{eq:Mass}
\end{eqnarray}
where ${\varepsilon}_0$ and $(j_0)_{\phi}$ are the energy density and 
angular momentum density derived from $S^0$. Primes denote the derivative 
with respect to $r$. Note that, at spatial infinity, mass and energy become 
identical with each other only when $N$ and $N^{\phi}$ should satisfy $N=1$ 
and $N^{\phi}=0$, respectively, and when we have a finite angular momentum 
there. As for the explicit expressions of $\varepsilon_0$ and $(j_0)_{\phi}$ 
we will discuss them later in connection with the divergence of the angular 
momentum.

The solutions for the metric components are given by
\begin{eqnarray}
	L&=&|\Lambda|^{1 \over 2}{{\rho^2-r_0^2}\over {\rho \rho'}},
	\label{eq:Lcomp}\\
	N&=&C|\Lambda|^{1\over 2} {{\rho^2-r_0^2}\over K },
	\label{eq:Ncomp}\\
	{\tilde N}^{\phi}&=&\varepsilon C|\Lambda|^{1\over 2} {{\rho^2-r_0^2}\over 
K^2},
	\label{eq:Nphicomp}
\end{eqnarray}
where ${\tilde N}^{\phi}$ is related to $N^{\phi}$ by
\begin{equation}
	{\tilde N}^{\phi} \equiv N^{\phi}+C_2/C_1.
	\label{eq:Relation}
\end{equation}
Here $\rho$ is a function of $r$ and related to $K$ as
\begin{equation}
	K^2=\rho^2+r_0^2 {\rm ln}|{{\rho^2-r_0^2}\over r_0^2}|+q,
	\label{eq:Rho}
\end{equation}
where $q$ is a constant and $r_0^2 \equiv 4\pi G C_1^2/|\Lambda|$. In the 
above equations, we can set $C=1$ without loss of generality and so we shall 
adopt it henceforth. We shall next determine the constant $C_1$ by the 
surface integral of the electric field $E_{\hat r}$. We identify it as the 
electric charge:$C_1=Q_e$. As a boundary condition, we assume that
\begin{equation}
	N^{\phi}(\infty)=0.
\label{eq:BC}
\end{equation}
When ${\rho}^2 \to \infty$ as $r \to \infty$, we can determine $C_2$ by 
using the boundary condition as
\begin{equation}
	C_2=\varepsilon Q_e|\Lambda|^{1 \over 2}.
	\label{eq:C2}
\end{equation}
The SD(ASD) equation (\ref{eq:SD}) renders Eqs.(\ref{eq:Elec}) and 
(\ref{eq:Mag}) into
\begin{equation}
	E_{\hat r}=\varepsilon {B_{\hat{~}}}={{Q_e} \over K}.
	\label{eq:SDEM}
\end{equation}
So far we have not specified the explicit form of the function $\rho(r)$. We 
shall call its explicit specification the coordinate condition. Of a lot of 
coordinate conditions we shall discuss some important cases.
\vskip 0.1 true in
\noindent
(i)$K=r$ case

 The most naive coordinate condition that $K=r$ together with $L=N$ leads to 
the line element of BTZ solution \cite{BTZ,BHTZ} and so we can not allow for 
neither electric nor magnetic field. This condition is too stringent to 
impose.
\vskip 0.1 true in
\noindent
(ii)$\rho=r$ case

We shall discuss the coordinate condition, which we adopted in the previous 
paper. The condition is to set $\rho=r$ in the solutions. This yields
\begin{equation}
	K^2=r^2+r_0^2 {\rm ln}|{{r^2-r_0^2}\over r_0^2}|+q.
	\label{eq:Cocond}
\end{equation}

Substituting the solutions and using the coordinate condition, we obtain the 
energy, the angular momentum and the mass as
\begin{eqnarray}
E&=&-2 |\Lambda|^{1\over2} {{r^2}\over{\sqrt{r^2+ r_0^2 {\rm 
ln}|{{r^2-r_0^2}\over{r_0^2}}|+q}}}-2\pi K\varepsilon_0,
\label{eq:Ersol}\\
	J&=&\varepsilon {{8 \pi G Q_e^2}\over|\Lambda|^{1\over2}}({\rm 
ln}|{{r^2-r_0^2}\over{r_0^2}}|+{q \over r_0^2})-2\pi K (j_0)_{\phi},
	\label{eq:Jsol}\\
	M&=&2 |\Lambda| \Bigl\{-r^2(r^2-r_0^2)+(r_0^2 {\rm 
ln}|{{r^2-r_0^2}\over{r_0^2}}|+q)(r_0^2+ r_0^2{\rm 
ln}|{{r^2-r_0^2}\over{r_0^2}}|+q)\Bigr\}/K^2
\nonumber\\
&-&2 \pi |\Lambda|^{1 \over 2} \{(r^2-r_0^2 ){\varepsilon}_0 + \varepsilon 
{{r_0^2 +r_0^2 {\rm ln}|{{r^2-r_0^2}\over{r_0^2}}| +q}\over K}(j_0)_{\phi} 
\}.
	\label{eq:Msol}
\end{eqnarray}
We note that the conserved charges, mass and angualr momentum, are not 
constants but $r$ dependent. The angular momentum has a logarithmic 
divergence as $r$ goes to infinity if we set $(j_0)_{\phi}=0$.  The mass has 
a quadratic divergence
 at spatial infinity while the energy has a linear divergence for 
$(j_0)_{\phi}=0$ and ${\varepsilon}_0=0$. Due to the logarithmic divergence 
of the angular momentum our solution could not be called the black hole 
solution.

In our previous paper we neglected the contribution from the boundary and 
set $(j_0)_{\phi}=0$ in obtaining the angular momentum. However in this 
paper we take the contribution from the boundary into account in determining 
the above quantities. As far as we do not have the criterion of defining the 
functional form of ${\varepsilon}_0$ and $(j_0)_{\phi}$, the above 
quantities can be put into any functional form by an appropriate choice of 
them. In order to avoid the arbitrariness, we adopt a criterion in the 
choice of them following the regularization procedure in the quntum field 
theory. In QED, the logarithmic divergence of mass operator is cancelled by 
the counter term leaving the finite part unchanged. We set the ansatz that 
${\varepsilon}_0$ and $(j_0)_{\phi}$ should be chosen so that they cancell 
the logarithmically divergent parts in angular mometum and mass, if any. In 
the present solution we have a logarithmically divergent angular momentum at 
spatial infinity if $(j_0)_{\phi}=0$. Following the criterion, we assume 
that $(j_0)_{\phi}$ is a function of $r$ which  eliminates the logarithmic 
divergence as a counter term.  We thus obtain the regularized angular 
momentum as
\begin{equation}
	J=-\varepsilon {{8 \pi G Q_e^2} \over |\Lambda|^{1\over2}}
\Bigl(1-{\rm ln}|{{r^2-r_0^2} \over {r^2-r_{\infty}^2}}|\Bigr),
	\label{eq:Regj}
\end{equation}
where $r_{\infty}$ is a constant. The regularized angular momentum is still 
$r$ dependent but free from the logarithmic divergence at spatial infinity. 
However the divergence on the horizon($r=r_0$) still exists. If we require 
that $r_0^2=r_{\infty}^2$, we can remove the divergence also and we are left 
with the constant angular momentum:
\begin{equation}
	J=-\varepsilon {{8 \pi G Q_e^2}\over|\Lambda|^{1\over2}},
	\label{eq:Regjfinal}
\end{equation}
which may be compared with the simultaneous regularization of the infrared 
and ultraviolet divergence in the quantum field theory. In a similar way we 
can cancel the divergences of energy and mass by adjusting 
$\varepsilon_0$.
With such regularized quantities the solution has a horizon and so we may 
call this the black hole solution.
\vskip 0.1 true in
\noindent
(iii)$g_{\phi \phi}(\infty) \not= r^2$ case

 In the case (ii) we found that the non-trivial $(j_0)_{\phi}$ as a counter 
term could lead to a black hole solution. We shall first clarify why we have 
such a logarithmic divergence of angular momentum for the $(j_0)_{\phi}=0$ 
case. When  we do not take the contribution from the boundary into account, 
we should discard the condition that $g_{\phi \phi}(=K^2)$ should approach 
the radial coordinate squared as far as we want the finite angular 
momentum.

 In order to make the point clearer,  we shall rewrite the 
Eqs.(\ref{eq:Lcomp})-(\ref{eq:Rho}). We introduce $x$ by
\begin{equation}
	x={{{\rho}^2-r_0^2}\over {r_0^2}}.
	\label{eq:Defx}
\end{equation}
Then the metric components read
\begin{eqnarray}
	L&=&2 |\Lambda|^{1\over 2}/({\rm ln}|x|)',
	\label{eq:Lx}\\
	N&=&|\Lambda|^{1\over 2} r_0^2 x/K,
	\label{eq:Nx}\\
	N^{\phi}&=&-\varepsilon |\Lambda|^{1\over 2}{{r_0^2}\over K^2}(1+{\rm 
ln}|x|+{q\over r_0^2}),
	\label{eq:Nphix}	
\end{eqnarray}
where $K^2$ is given by
\begin{equation}
	K^2=r_0^2(1+x+{\rm ln}|x|+{q\over {r_0^2}} ).
      	\label{eq:Kx}
\end{equation}
 To impose the boundary condition that $g_{\phi \phi}(=K^2)$ should approach 
 $r^2$ as $r$ becomes infinite implies that $x$ approaches infinity as is 
seen from this equation. Then this causes the logarithmic divergence of the 
angular momentum, which behaves as ${\rm ln} x$ at infinity, in the 
$(j_0)_{\phi}=0$ case. Therefore we need to discard the condition that 
$g_{\phi \phi}$ should approach the radial coordinate squared if we want the 
finite angular momentum with zero $(j_0)_{\phi}$. The origin of the 
divergence of angular momentum can be considered to be the nonzero matter 
momentum density:``Poynting pseudovector'' of the electromagnetic fields 
(19).

 In order to obtain the black hole solution without any divergence of 
angular momentum, we require firstly that there should be a horizon and 
secondly that the angular momentum is finite all through the region from the 
horizon to infinity. The finiteness of mass shall be checked after finding 
the solutions satisfying the above conditions. In order to realize the above 
requirements we should find the coordinate condition $x=g(r)$ which satisfy 

\begin{eqnarray*}
(I)\hspace{5mm}L&\sim&g(a)/g'(a)=0,\hspace{5mm} {\rm at} \hspace{5mm} 
r=a,\\
(II)\hspace{5mm}J&\sim&|{\rm ln} g(r)|< \infty,\hspace{5mm} {\rm for} 
\hspace{5mm} a \leq r < \infty.
\end{eqnarray*}
Here $r=a$ is the position of horizon. The previous solution does not match 
the conditions. Since there is an infinite class of solutions satisfying the 
conditions, we can not exhaust them all, but we shall illustrate one of the 
class of solutions. Consider $g(r)$ defined by
\begin{equation}
	{g(r)\over g'(r)}=A \sqrt{r-a}e^{B(r-a)},
\label{eq:Deqg}
\end{equation}
where $A$ and $B$ are positive constants. This is easily integrated to 
give
\begin{equation}
	g(r)=e^{-{2 \over {A\sqrt B}}{\rm Erfc}\left(\sqrt{B(r-a)}\right)}.
\label{eq:Gsol}
\end{equation}
Here ${\rm Erfc}(x)$ is the error function defined by
\begin{equation}
	{\rm Erfc}(x)=\int_x^{\infty}e^{-t^2}dt,
\label{eq:Erf}
\end{equation}
which is a monotonically decreasing function varying from $\sqrt{\pi}/2$ to 
0 as $x$ changes from 0 to $\infty$. By using the solution the coordinate 
condition is explicitly given by
\begin{equation}
	{\rho}^2=r_0^2 \Bigl\{1+e^{-{2\over{A{\sqrt B}}}{\rm 
Erfc}\left(\sqrt{B(r-a)}\right)} \Bigr\}.
\label{eq:Solrho}
\end{equation}
By using the boundary condition(\ref{eq:BC}), we obtain $q/r_0^2=-1$. The 
metric components are now written as
\begin{eqnarray}
	L &=&2 |\Lambda|^{1 \over 2} A \sqrt{r-a} e^{B(r-a)},
	\label{eq:Lsol}\\
	N &=& |\Lambda|^{1 \over 2} r_0^2 g(r) /K,
	\label{eq:Nsol}\\
	N^{\phi} &=& -\varepsilon |\Lambda|^{1 \over 2} r_0^2{{\rm ln}|g(r)| \over 
K^2},
	\label{eq:Nphisol}
\end{eqnarray}
where $K^2$ is given by
\begin{equation}
K^2 = r_0^2(g(r)+{\rm ln}|g(r)|).
\label{eq:Kbyr}
\end{equation}

The solution shows a peculiar topology of the spacetime. $K^2$ approaches 
$r_0^2$ as $r$ goes to infinity. As is seen from Eq.(\ref{eq:Lsol}) and 
Eq.(\ref{eq:Nsol}), $L^2$ vanishes at $r=a$, but $N^2$ is finite there. The 
angular momentum reads
\begin{equation}
J=-\varepsilon {{8 \pi G Q_e^2}\over{|\Lambda|^{1\over2}}}\Bigl\{{2\over 
{A\sqrt B}}{\rm Erfc}\left(\sqrt{B(r-a)}\right)+1\Bigr\},
\label{eq:FiniteJ}
\end{equation}
where we have set $(j_0)_{\phi}=0$. As far as the constants $A$ and $B$ 
satisfy the condition:
\begin{equation}
e^{-{1 \over A}\sqrt{{\pi \over B}}}-{1 \over A}\sqrt{{\pi \over B}} \geq 
0,
\label{eq:Ieqlty}
\end{equation}
$K^2$ is kept positve for $a \leq r < \infty$ and the metric components are 
finite there. Though a further physical study is necessary till we can 
conclude that this is really a black hole solution, the peculiar spacetime 
is intriguing enough for further investigation.

We have seen how the logarithmic divergence of the angular momentum is 
regularized by taking the contribution from the boundary into account. If 
the relation
$r_0=r_{\infty}$ between two parameters $r_0$ and $r_{\infty}$ should hold, 
we have the constant angular momentum and so we have a charged black hole 
solution. Even when we neglect the effect of the boundary, there is a 
peculiar solution with a finite angular momentum. We showed the close 
relation between the behavior of $g_{\phi \phi}$ and angular momentum at 
spatial infinity. The finiteness of the angular momentum is related to the 
fact that $g_{\phi \phi}$ approaches constant at the spatial infinity. We 
shall discuss the physical aspects of these solutions in the forthcoming 
paper.

\end{document}